# Magnetocrystalline anisotropy of the easy-plane metallic antiferromagnet Fe$_2$As

Kexin Yang,[1,2] Kisung Kang,[2,3] Zhu Diao,[2,4,5,†] Manohar H. Karigerasi,[2,3] Daniel P. Shoemaker,[2,3] André Schleife,[2,6] David G. Cahill [1,2,3,*]

[1]Department of Physics, University of Illinois at Urbana-Champaign, Urbana, Illinois 61801, USA

[2]Materials Research Laboratory, University of Illinois at Urbana-Champaign, Urbana, Illinois 61801, USA

[3]Materials Science and Engineering, University of Illinois at Urbana-Champaign, Urbana, Illinois 61801, USA

[4]Department of Physics, AlbaNova University Center, Stockholm University, SE-106 91 Stockholm, Sweden

[5]Department of Physics, Florida A&M University, Tallahassee, Florida 32307, USA

[6]National Center for Supercomputing Applications, University of Illinois at Urbana-Champaign, Urbana, Illinois 61801, USA

**ABSTRACT**

Magnetocrystalline anisotropy is a fundamental property of magnetic materials that determines the dynamics of magnetic precession, the frequency of spin waves, the thermal stability of magnetic domains, and the efficiency of spintronic devices. We combine torque magnetometry and density functional theory calculations to determine the magnetocrystalline anisotropy of the metallic antiferromagnet Fe$_2$As. Fe$_2$As has a tetragonal crystal structure with the Néel vector lying in the (001) plane. We report that the four-fold magnetocrystalline anisotropy in the (001)-plane of Fe$_2$As is extremely small, $K_{22} = -150$ J/m$^3$ at $T = 4$ K, much smaller than perpendicular magnetic anisotropy of ferromagnetic structure widely used in spintronics device. $K_{22}$ is strongly temperature dependent and close to zero at $T > 150$ K. The anisotropy $K_1$ in the (010) plane is too large to be measured by torque magnetometry and we determine $K_1 = -830$ kJ/m$^3$ using first-principles density functional theory. Our simulations show that the contribution to the anisotropy from classical magnetic dipole-dipole interactions is comparable to the contribution from spin-orbit coupling. The calculated four-fold anisotropy in the (001) plane $K_{22}$ ranges from $-290$ J/m$^3$ to 280 J/m$^3$, the same order of magnitude as the measured value. We used $K_1$ from theory to predict the frequency and polarization of the lowest frequency antiferromagnetic resonance mode and find that the mode is linearly polarized in the (001)-plane with $f = 670$ GHz.

[*] d-cahill@illinois.edu
[†] zhu.diao@famu.edu

**INTRODUCTION**

Antiferromagnets (AFs) have potential advantages over ferromagnets for spintronic devices. Collinear AFs are relatively insensitive to external fields because the net magnetization is zero. AFs typically have much higher antiferromagnetic resonance (AFMR) frequency than ferromagnets and therefore processional switching can occur in AFs at a faster rate than in ferromagnets.

The recent discovery of *electrical* manipulation and detection of spin configurations in metallic AFs has led to a rapidly expanding scientific literature on this class of magnetic materials. Tetragonal crystals with easy-plane magnetic anisotropy are preferred because the two degenerate orientations of the Néel vector can store binary information. In crystals with globally centrosymmetric but locally non-centrosymmetric magnetic structures—e.g., CuMnAs and $Mn_2Au$—an electrical current exerts a torque on the Néel vector and the domain structure can potentially be switched electrically [1][2][3][4].

A small value of the in-plane magnetocrystalline anisotropy facilitates electrical switching of the domain orientation since a smaller torque is needed to overcome the energetic barrier that separates the two orientations. Thermal stability of the domain requires, however, a large value of the in-plane anisotropy. The Néel-Arrhenius law provides an estimate of the rate of thermal fluctuations of a single domain[5]:

$$\frac{1}{\tau} = f_0 \exp\left(-\frac{\Delta E}{k_B T}\right) \quad , \tag{1}$$

where $\tau$ is the average time between thermally-activated changes in the direction of the magnetization, $f_0$ is the resonance frequency, $\Delta E$ is the energy barrier between two degenerate magnetic states and $k_B T$ is the thermal energy. $\Delta E$ is given by the product of an anisotropy parameter $K$ and the volume of the domain $V$; $\Delta E = KV$. Stable data storage typically requires $\Delta E / k_B T > 40$ to meet the criteria that data must be retained for 10 years [6].

For the media of conventional hard drives, the anisotropy parameter $K$ is controlled by the perpendicular magnetocrystalline anisotropy of ordered intermetallic alloys. In the emerging technology of magnetic random access memory (MRAM), $K$ is controlled by the interfacial magnetic anisotropy of a ferromagnetic layer adjacent to the oxide barrier in a magnetic tunnel junction. The perpendicular magnetic anisotropy $K_1$ of MRAM materials is typically $10^6 < K_1 < 10^7$ J m$^{-3}$ [7].

Magnetocrystalline anisotropy $E_{ani}$ is described by a phenomenological expansion of the energy as a function of direction cosines for the orientation of magnetization of a ferromagnet or the sublattice magnetization of an antiferromagnet (AF). For a tetragonal crystal, the expansion to fourth order gives 3 coefficients $K_1$, $K_2$ and $K_{22}$ [8]. $K_1$ is a second order coefficient; $K_2$ and $K_{22}$ are fourth order coefficients.

$$E_{ani}/V = K_1 \sin^2\phi + K_2 \sin^4\phi + K_{22} \sin^4\phi \cos(4\theta) \quad , \qquad (2)$$

where $\phi$ is the angle of the magnetization relative to the <001> direction and $\theta$ is the angle of the magnetization relative to the <100> direction (Fig. 2). The coefficient $K_1$ describes the two-fold anisotropy in (010) plane and $K_2$ represents the higher order four-fold symmetry of the (010) plane. Because the effect of $K_2$ is usually much smaller than $K_1$, $K_2$ will be neglected in the following discussion. A crystal with an easy-plane anisotropy is described by $K_1 < 0$. The coefficient $K_{22}$ describes the four-fold anisotropy of the (001) plane and determines the thermal stability of an easy-plane domain structure.

An external magnetic field applied to an antiferromagnet (AF) produces a small induced magnetic moment. The induced moment is small because tilting of the orientation of sublattice magnetization is constrained by strong exchange interaction between the magnetic sublattices that favors a parallel alignment of the sublattices. In general, however, the induced magnetic moment is not parallel to the applied field because magnetocrystalline anisotropy favors an orientation of the sublattice magnetization along an easy axis [9][10].

The lack of alignment between the induced moment **m** and the applied field **B** produces a macroscopic torque on the sample, $\tau = m \times B$. A torque magnetometer measures this torque. Data for the torque as a function of applied field is sensitive to magnetocrystalline anisotropy as long as the anisotropy is neither too small nor too large. If the anisotropy is small, then the angle between **m** and **B** is small and the torque becomes difficult to detect. If the anisotropy is large, then the direction of **m** is fixed with respect to the crystallographic axis and the torque does not provide information about the magnitude of the anisotropy. We can measure the in-plane four-fold anisotropy of a mm-size bulk crystal of $Fe_2As$ by torque magnetometry but the out-of-plane two-fold anisotropy is not accessible to this technique because the external field is too small compare to anisotropy field to extract information about the anisotropy in the out-of-plane direction. We instead employ first-principles calculations based on density functional theory to determine $K_1$.

When magnetic energy is larger than the anisotropy energy, the amplitude of the torque in the (001)-plane saturates and the four-fold magnetic anisotropy, $K_{22}$ can be directly determined from the amplitude of the torque. We measured three samples extracted from the same growth run, and the $K_{22}$ value of all three samples is comparable to -150 $J/m^3$ at 4 K. The magnitude of $K_{22}$ drops quickly as temperature increases and reaches a small value above 150 K. The temperature dependence of magnetic anisotropy for antiferromagnets is similar to ferromagnets, following a power law of sublattice magnetization [11][12][13].

Strikingly, torque data for the applied field rotating in the (010)-plane reveal the motion of domain walls. An applied field in the (010) plane of 1 T is sufficient to orient the Néel vector fully perpendicular to the applied field. Domain wall motion occurs even at $T = 4$ K and, therefore, is not thermally activated.

In the final section, we derive the lowest-frequency, zone-center AFMR frequency for easy-plane AFs, $\omega = |\gamma|\sqrt{2H_E(H_{22} - H_1)} \approx |\gamma|\sqrt{-2H_E H_1}$, where $\gamma$ is the gyromagnetic ratio, and $H_E$, $H_1$, $H_{22}$ are the exchange field, out-of-plane anisotropy field and in-plane anisotropy field, respectively. The anisotropy fields are calculated with anisotropy energy and sublattice magnetization: $H_1 = K_1 / M$ and $H_{22} = K_{22} / M$. With $K_1$ calculated by DFT as $K_1$ = -830 $kJ/m^3$, the AFMR frequency is $f = 670$ GHz at 4 K.

**METHODS**

$Fe_2As$ crystallizes in the $Cu_2Sb$ tetragonal crystal structure. Based on the corresponding magnetic symmetry (mmm1' magnetic point group), the Néel vector of $Fe_2As$ has two degenerate orientations in the (001)-plane [14][15][16].

The $Fe_2As$ crystal was synthesized by mixing Fe and As powders in a 1.95:1 ratio and vacuum sealing inside a quartz tube. The vacuum tube was heated at 1 °C/min up to 600 °C and held for 6 hours in a furnace. The temperature was then ramped to 975 °C at 1 °C/min and held for 1 hour before cooling down slowly to 900 °C at 1 °C/min. Finally, the quartz tube was kept at 900 °C for 1 hour and allowed to cool down to room temperature in the furnace at 10 °C/min. We obtained a large silver-hued crystal ingot of $Fe_2As$ and it easily detached from the quartz tube. Part of the ingot was crushed into powder for powder XRD characterization and the data showed phase pure $Fe_2As$. But the sample is slightly off-stochiometry as described in reference [17]. The remaining portion of the ingot was then fractured and the fractured surface revealed a smooth facet. Laue diffraction was carried out after polishing this fractured surface. A four-fold symmetry pattern was observed indicating the fractured surface is the (001) plane.

We used a wire saw to cut the sample into smaller pieces for magnetic property characterization and torque measurements. One of the pieces was measured on the superconducting quantum interference device vibrating sample magnetometer (SQUID-VSM, see below), the other three pieces were used in torque magnetometry measurements. We name these three samples measured by torque magnetometry sample A, sample B, sample C and the one for SQUID-VSM is named sample D.

The temperature-dependent magnetic susceptibility was measured with a SQUID-VSM in a Quantum Design Magnetic Properties Measurement System (MPMS). The susceptibility of the sample was measured while cooling from 398 K to 4 K in a 10 mT field.

Torque measurements were performed in a Quantum Design Physical Property Measurement System (PPMS). We mounted the sample on a standard torque sensor chip, P109A from Quantum Design with a sensitivity of $1 \times 10^{-9}$ N·m. The PPMS horizontal sample rotator was used to control the angle between the crystal and the applied field. During the measurement, the external field rotated in either the (010) plane or the (001) plane, while the field-induced moment resided in the same plane as the rotating applied field. We detected the torque component, $m \times B$, that is perpendicular to this plane.

We performed first-principles DFT simulations using the Vienna *Ab Initio* Simulation Package (VASP)[18][19], to calculate the two-fold anisotropy $K_1$ and obtain an estimate for the four-fold anisotropy $K_{22}$. The projector augmented wave (PAW) method [20] is used to describe the electron-ion interaction. Kohn-Sham states are expanded into plane waves up to a kinetic energy cutoff of 600 eV. The Brillouin zone is sampled by a $21 \times 21 \times 7$ Monkhorst-Pack [21] (MP) **k**-point grid and the total energy is converged self-consistently to within $10^{-9}$ eV. The local density approximation (LDA) [22] and the generalized-gradient approximation developed by Perdew, Burke, and Ernzerhof (PBE) [23] are used to describe the exchange-correlation energy function, and results from the two different computational strategies are compared.

Achieving the extremely high accuracy for total energies that is required to compute the (001) plane magnetocrystalline anisotropy that is on the order of $\mu$eV per magnetic unit cell is numerically challenging; the required convergence parameters render it computationally too expensive to perform such calculations fully self-consistently. Instead, we use the convergence parameters quoted above to compute Kohn-Sham states, electron density, and relaxed atomic geometries for collinear magnetism and take non-collinear magnetism and spin-orbit coupling [24] into account without self-consistency of the Hamiltonian, as described in Ref. [25].

**RESULTS AND DISCUSSION**

**1. Magnetic susceptibility and domain wall motion**

We use data for the magnetic susceptibility as input for modeling the torque magnetometry data and to provide insight into the reorientation of antiferromagnetic domains in an external magnetic field. Figure 1(a) summarizes the results for the magnetic susceptibility in the limit of small field. We fixed the applied field at 10 mT, along the <100> or <001> direction, measured the induced magnetic moment while cooling from $T = 398$ K, and calculated the susceptibility, $\chi = M/H$, where $M$ is the magnetization. The measured susceptibility is similar to that in a prior report [14].

When the applied field is along an easy axis, we must take domain wall motion into account. We assume that a 10 mT external field is too weak to significantly affect the domain structure. We further assume that the magnetic moment generated by an applied field along the <100> direction (the *a*-axis of the crystal) has equal contributions from two types of domains that we label as D1 (Néel vector along <100>) and D2 (Néel vector along <010>) as illustrated in Fig. 2. For an applied field in the (001)-plane, we define the susceptibility parallel to the Néel vector as $\chi_\parallel$ while that perpendicular to the Néel vector as $\chi'_\perp$. On the other hand, the susceptibility for an applied field in the <001> direction is defined as $\chi_\perp$. We expect $\chi'_\perp$ and $\chi_\perp$ to be similar but due to the tetragonal symmetry of the crystal structure, $\chi'_\perp$ and $\chi_\perp$ are not necessarily equal. We show below that the difference between $\chi_\perp$ and $\chi'_\perp$ is less than 5%.

Measurements of the magnetization as a function of field, see Fig. 1(b), show that $\chi_c$ is constant for $H$ applied along the *c*-axis. For $H$ along the *a*-axis, $\chi_a$ increases with field at low field, and is approximately constant for an applied field > 1 T. We attribute the field dependence of $\chi_a$ to domain wall motion and the consequent evolution of the populations of domains with Néel vectors parallel and perpendicular to the applied field.

The populations of the two degenerate domains can be estimated from the *M* vs *H* curve in Fig. 1(a) by expressing the field-induced magnetization as $M_a = \sigma_\perp \chi'_\perp H + \sigma_\parallel \chi_\parallel H$ and $M_c = \chi_\perp H$, where $\sigma_\perp$ and $\sigma_\parallel$ are the normalized domain fraction perpendicular and parallel to the *a*-axis, respectively, and $\sigma_\perp + \sigma_\parallel = 1$. We made three assumptions: (1) $\sigma_\perp = \sigma_\parallel = 0.5$ at zero field; (2) $\sigma_\perp \approx 1$ at high field; and (3) domain wall motion is reversible. The field-dependent distribution of domains parallel and perpendicular to the external field along the *a*-axis at $T = 4$ K is shown in Fig. 1(c) and are treated as free parameters; we find $\chi_\parallel = 0.008$ and $\chi_\perp$ anisotropy in the= 0.018. For an ideal collinear antiferromagnet, we expect $\chi_\parallel = 0$ at low temperatures [9]. This is not what we observed in our measurements. The reason is that there is a background contribution to the magnetic susceptibility that we do not yet understand. We assume that the background susceptibility is isotropic.

## 2. Torque magnetometry

The field-induced torque is the cross product of the field-induced magnetic moment and the applied field, $\tau = m \times B$. The direction of the induced magnetic moment $m$ is given by the minimum in the total energy: $E_{tot} = E_m + E_{ani} + E_{ex}$, where $E_m$ is the magnetic energy; $E_{ani}$ is the magnetocrystalline anisotropy energy; and $E_{ex}$ is the exchange energy that couples the two sublattices. We refer to the condition $E_m \ll E_{ani}$ as the low field limit and the condition $E_m \geq E_{ani}$ as the intermediate field regime[26]. We ignore a separate $E_{ex}$ term when we analyze the torque data in the (001)-plane because we assume that the exchange interaction stays the same and can be represented by susceptibility. For torque data in the (010) plane, our analysis is based on the anisotropy in the susceptibility, $\chi_\perp - \chi_\parallel$, which is also related to the strength of the exchange interaction.

Fig. 2(a) shows the experimental geometry when the applied field is rotating in the (010)-plane. $\xi$ is the angle between the $c$-axis and the applied field. The induced moments are also in the (010)-plane. Thus, the torque is along the <010> direction. In the low temperature limit, $\chi_\parallel = 0$ if the isotropic background is ignored; the induced moment of domain D1 is therefore along <001> and the induced moment of D2 lies in the (010) plane between <001> and <100>. The direction of the induced moment of D2 is determined by $\chi_\perp$ and $\chi'_\perp$.

Fig. 2(b) shows the experimental geometry when the applied field is rotating in the (001)-plane. In this case, because of the relatively small magnetocrystalline anisotropy, the tilt of the sublattice magnetization away from the crystal axes is significant. $\psi$ is the angle between the external field and the crystal axes; $\theta_1$, $\theta_2$ are the angles between the directions of the sublattice magnetization of domains D1, D2 and the crystal axes, respectively ($\theta_1$ and $\theta_2$ are not necessarily equal).

## 2.1 Torque magnetometry in the (010)-plane in the low field limit

The torque is zero when the applied field is along the easy or hard axis of a sample, because the induced magnetization is in the same direction as the field. Here, we refer to the easy axis as the lowest energy orientation of the Néel vector, and define orientations of the hard axis as perpendicular to the easy axis. When the applied field is oriented away from an easy or hard axis, the direction of the induced magnetization shifts toward a hard axis because $\chi_\perp > \chi_\parallel$. In an AF, the slope of the torque as a function of field orientation has opposite signs when the field passes through the orientation of an easy axis and when it passes through the orientation of a hard axis. In our sign convention, torque with a negative slope as a function of angle indicates a hard axis; torque with a positive slope as a function of angle indicates an easy axis. In a single magnetic domain of $Fe_2As$, there are two hard axes: the $c$-axis perpendicular to the (001) plane and the axis perpendicular to the Néel vector in the (001) plane.

Torque data at 4 K with the field rotating in the (010)-plane are shown in Fig. 3(a). The slope when the field is along the $a$-axis ($\xi = 90°$) is positive at $B = 0.5$ T and changes to negative at $B > 0.5$ T. Therefore, when the 0.5 T field is oriented along the $a$-axis, the $a$-axis is an easy axis but when $B$ is greater than 0.5 T, the $a$-axis becomes a hard axis. This interpretation is consistent with the analysis of the domain distribution discussed above and displayed in Fig. 1(c). When the applied field along the $a$-axis is larger than 1 T, the majority of domains are in the D2 configuration and the $a$-axis becomes a hard axis.

The slope of torque data when the applied field is along the $c$-axis ($\xi = 0°$) is negative because the $c$-axis is a hard axis. However, when $\xi = 10°$ and $B > 0.5$ T, the sign of the torque changes abruptly. This dramatic change in the sign of the torque is periodic; the periodicity indicates that domain wall motion is reversible. When the applied field is aligned along the $c$-axis, the populations of domain D1 and D2 are equal. As the field rotates away from the $c$-axis, the projection of the applied field in the (001)-plane changes the domain distribution as described by Fig. 1(c). When the field returns to the $c$-axis, the populations of domain D1 and domain D2 become equal again.

To model the torque data, we first calculate the direction and magnitude of the induced magnetization $M$ by describing the susceptibility as a tensor $M_i = \sum_j \chi_{ij} H_j$ [8]. As shown in Fig. 2(a), there are two degenerate domains with their Néel vectors perpendicular to each other. For domains of type D1, the Néel vector is along the $a$-axis and the susceptibility tensor is

$$\chi_{D1} = \begin{pmatrix} \chi_\parallel & 0 & 0 \\ 0 & \chi'_\perp & 0 \\ 0 & 0 & \chi_\perp \end{pmatrix}. \tag{3}$$

For domains of type D2, the Néel vector is along the $b$-axis and the susceptibility tensor is

$$\chi_{D2} = \begin{pmatrix} \chi'_\perp & 0 & 0 \\ 0 & \chi_\parallel & 0 \\ 0 & 0 & \chi_\perp \end{pmatrix}. \tag{4}$$

The external field in the (010)-plane is $H = H_0 (\sin(\xi) \quad 0 \quad \cos(\xi))^T$, where $\xi$ is the angle between the $c$-axis and the external field as depicted in Fig. 2(a). We consider the effect of the projection of the applied field along the $a$-axis $H_0 \sin(\xi)$ on the domain distribution as described by the data of Fig. 1(c). The torque signal of two types of domains are $\tau_{D1}/V = \sigma_{D1} \chi_{D1} H \times B$ and $\tau_{D2}/V = \sigma_{D2} \chi_{D2} H \times B$, while the total torque is the sum $\tau_{D1} + \tau_{D2}$. To evaluate this model, we use the measured magnetic susceptibility as shown in Fig. 1. The free parameters are $\chi'_\perp / \chi_\perp$ and $\chi_\parallel$.

Fig. 3(b) and (c) show the calculated values of $\tau_{D1}$ and $\tau_{D2}$; the solid lines in Fig. 3(a) are the summation of the two. The good correspondence between the model and the data supports our assertion that domain wall motion is reversible.

The difference in the sign of $\tau_{D1}$ and $\tau_{D2}$ contributes to the abrupt change in the torque signal near an angle of 10°. For D1 domains, when the applied field is rotating in the (010)-plane, the induced moment is always close to the $c$-axis because $\chi_\parallel$ is small. For D2 domains, the field-induced magnetization is $M = H_0 (\chi'_\perp \sin(\xi) \quad 0 \quad \chi_\perp \cos(\xi))^T$. If $\chi'_\perp = \chi_\perp$, the induced magnetization $M = \chi_\perp H_0 \begin{pmatrix} \sin(\xi) & 0 & \cos(\xi) \end{pmatrix}^T$, would be parallel to the applied field $H$. This would infer that there is no torque signal from D2 domains and the total torque signal would be generated only by D1 domains (Fig. 3(b)).

However, the total torque signal we observe is obviously different from what is depicted in Fig. 3(b) (D1 domains only). The torque signal resembles a combination of two domains, hence, we can conclude $\chi'_\perp \neq \chi_\perp$. On the other hand, the dramatic change in the sign of the total torque signal at $\xi = 10°$ and $T = 4$ K indicates that $\tau_{D1}$ and $\tau_{D2}$ have opposite signs. Since the induced moment of D1 domains is along the $c$-axis, the induced moment of D2 domains must be between $B$ and the $a$-axis.

The difference between $\chi_\perp$ and $\chi'_\perp$ is also observed in the dependence of $M$ on $H$ (Fig. 1(b)). Figure 4(a) shows the difference between $M_c$ and $M_a$ as a function of applied field and temperature. The bump of $M_c - M_a$ at room temperature and below is due to domain wall motion; the difference at high fields is a result of $\chi_\perp - \chi'_\perp$.

The magnetization of D2 domains in the (010)-plane experiences an anisotropic environment that originates from the difference of the $a$- and $c$-axis of the crystal. By fitting the model to the torque data, we determine $\chi'_\perp - \chi_\perp$ as a function of temperature (Fig. 4(b)). We observe a change in sign of $\chi_\perp - \chi'_\perp$ near 200 K that is consistent with the anisotropy in the dependence of $M$ on $H$ (Fig. 4(a)). This change in sign indicates that the field-induced magnetization of D2 domains is between the applied magnetic field and the $a$-axis below 200 K, and between the applied magnetic field and the $c$-axis above 200 K. The anisotropy in the perpendicular susceptibility, $\chi_\perp - \chi'_\perp$, is always small compared to its absolute value: $|\chi_\perp - \chi'_\perp|/\chi_\perp \leq 0.05$.

## 2.2 Torque magnetometry in the (001)-plane under intermediate field

Figure 5(a) shows torque data acquired with the applied field rotating in the (001) plane. As expected, the data show four-fold symmetry. We attribute the small two-fold symmetry to a background that comes from misalignment of the sample. (The $c$-axis is not precisely perpendicular to the field direction.) The amplitude of the four-fold contribution to the torque as a function of applied field of three samples from the same growth run is plotted in Fig. 5(b).

To quantify the magnetic anisotropy in the (001)-plane of Fe$_2$As, we analyze the torque data by minimizing the total energy for D1 domains and D2 domains, respectively, then add $\tau_{D1}$ and $\tau_{D2}$ together to compare it to the data. When $E_m$ is comparable to or larger than $E_{ani}$ ($E_m \geq E_{ani}^{(001)}$), Néel vectors start tilting away from crystal axes as shown in Fig. 2(b). We assume that the two sublattice magnetizations are approximately parallel to each other, so the exchange interaction is considered in magnetic energy. With the applied field rotating in the (001)-plane of Fe$_2$As, the total energy is

$$E_{tot}/V = -\frac{1}{2}H(\psi)\chi_{tot}(\psi,\theta)H^T(\psi) + K_{22}\cos(4\theta) + K_1 + K_2 , \tag{5}$$

where $\chi_{tot} = \sigma_{D1}\chi_{D1}$, $\theta = \theta_1$ for D1 domains and $\chi_{tot} = \sigma_{D2}\chi_{D2}$, $\theta = \theta_2$ for D2 domains.

To obtain an accurate value of $E_m$, we rotate the susceptibility tensor together with the Néel vector $\chi^R(\theta) = R(\theta)\chi R^T(\theta)$ [8], where $\theta$ is the angle between the Néel vector and the crystal axis, and $R(\theta)$ is the rotation matrix:

$$R(\theta) = \begin{pmatrix} \cos(\theta) & -\sin(\theta) & 0 \\ \sin(\theta) & \cos(\theta) & 0 \\ 0 & 0 & 1 \end{pmatrix} \tag{6}$$

During the rotation, the field component along $M_{D1}$ and $M_{D2}$ result in a change in the populations of D1 and D2 domains. Thus, $\sigma_{D1}$ and $\sigma_{D2}$ are determined by the angle between the applied field and the spin axis, $(\theta + \psi)$.

The analogous behavior of a uniaxial antiferromagnet (AF) [10] provides a point of comparison. In a uniaxial AF, the critical field for the spin-flop transition is $H_c = \sqrt{2(K_1 + K_2)/(\chi_\perp - \chi_\parallel)}$. For Fe$_2$As, as shown in Fig. 1(c) and Fig. 5(b), we do not observe a sudden change in the domain populations that would be characteristic of a spin-flop transition. Furthermore, in a perfect crystal that is free from disorder, the single domain structure created by an applied field would persist after the field is removed. (In ferromagnets, domains form to reduce the contribution of the magnetic energy of stray fields to the total energy. In AFs, this driving force for domain formation is absent.) We attribute gradual and reversible domain movement in Fe$_2$As to random strain fields created by static disorder in the crystal that create local variations in the anisotropy energy.

The midpoint of the change in the populations of the D1 and D2 domains as a function of field is, however, close to what is expected for the characteristic field of a spin-flop transition of an ideal single domain easy plane antiferromagnetic crystal. The total energy of D2 when $\xi = 90°$ is [8]

$$\frac{E_{tot}}{V} = K_1 + K_2 + K_{22}\cos(4\theta_2) + \frac{1}{2}(\chi_\perp - \chi_\parallel)H^2\cos^2(\psi + \theta_2) - \frac{1}{2}\chi_\perp H^2. \qquad (7)$$

Both anisotropy and magnetic energy are function of $\theta_2$ but with different periodicity. For an applied field along the <010> direction ($\psi = 0°$), $E_{tot}(\theta_2 = 90°)$ is always a global minimum while $E_{tot}(\theta_2 = 0°)$ is at a local minimum under low fields and it becomes the global maximum when the applied field is larger than a characteristic field $H_c$. We can derive this critical field from $\theta_{max} = \frac{1}{2}\arccos\frac{(\chi_\perp - \chi_\parallel)H^2}{16|K_{22}|}$ in which $\theta_{max}$ represents the spin orientation when total energy is at global maximum under a known applied field. When $H \geq H_c$, $\theta_{max}(H \geq H_c) = 0°$. Therefore, $\mu_0 H_c = \sqrt{16|K_{22}|/(\chi_\perp - \chi_\parallel)} = 560$ mT using the value of $K_{22} = -150$ J/m$^3$ from our torque magnetometry measurement as described below.

We attribute our observations that domains begin to move in an applied field smaller than $H_c$ and domain wall motion is not complete until the applied field is larger than $H_c$ to an inhomogeneous distribution of local values of the magnetocrystalline anisotropy. We speculate that random strains in the crystal caused by point and extended defects create a relatively broad distribution of anisotropy at different locations in the sample. When the external field is removed, the random strain field controls the energy of the domain orientation and the volume fraction of domains recovers its initial states.

The torque induced by the applied field is the derivative of the total energy $\tau = dE_{tot}/d\psi$. In the intermediate field regime, i.e., $E_m \geq E_{ani}$ or $H \geq H_c$, we assume that the sample is a single domain and the sublattice magnetization is always nearly perpendicular to the applied field, i.e., $\theta + \psi \approx \pi/2$. The "intermediate field" regime refers to an applied field larger than the characteristic field, but not large enough to significantly change the exchange interaction. An important assumption here is that the tilt of the two sublattice magnetizations in the external field is small enough to be neglected. With this approximation, the magnetic energy is nearly independent of $\theta$ and $\psi$, $E_m \approx -\frac{1}{2}\chi_\perp H_0^2$. Then, the torque can be easily related to the anisotropy: $\tau = \frac{dE_{tot}}{d\psi} \approx \frac{dE_{ani}}{d\psi} = -4K_{22}\sin(4\psi)$, where $\tau$ no longer depends on the magnitude of the applied field.

A graduation reorientation of domains of an easy-plane antiferromagnet as a function of applied field was also recently observed in 50 nm thick CuMnAs epitaxial layers grown on GaP [27]. (CuMnAs and Fe$_2$As have essentially the same crystal structure with Cu and Mn atoms in CuMnAs occupying the same lattice sites as the two crystallographically distinct Fe atoms in Fe$_2$As.) The strength of the field needed to reorient antiferromagnetic domains in CuMnAs epitaxial layers is similar to what we observe in Fe$_2$As bulk crystals. Thinner, 10 nm thick, CuMnAs layers show a pronounced in-plane uniaxial anisotropy and a more abrupt transition in domain structure as a function of field than 50 nm thick layers. X-ray magnetic linear dichroism (XMLD) measurements of CuMnAs epitaxial layers reveals that the domain reorientation is not fully reversible and hysteretic for fields less than 2 T. X-ray photoelectron microscopy (XPEEM) images acquired after applying 7 T fields in orthogonal directions also show that the domain structure does not revert to a fixed configuration in zero field.

Based on our observation in Fig 5(b), as field increases, $\tau$ increases quickly then saturates. At higher fields, $\tau$ is slightly smaller than the saturation value, rather than staying the same until 9 T. At higher fields, the tilting of spins caused by the external field cannot be neglected, so exchange interaction is no longer a constant. In our model, however, the torque stays the same after saturation based on our assumption of constant susceptibility and exchange interaction. This assumption is no longer valid in higher field. While induced magnetization is smaller than $\chi H$, the torque is also smaller than the saturation value.

As our model predicts, the experimentally measured torque amplitude saturates as the applied field approaches 1 T for sample B, and 3 T for samples A and C. Hence, it is safe to select 1 T and 3 T as the "intermediate field" regime for sample B and for samples A and C, respectively. The measured $K_{22}$ value of sample A is -150 J/m$^3$. The field-dependence of $K_{22}$ in all three samples follow the same trend; however, individual data points do not overlap perfectly. We attribute this discrepancy to variations in the defect microstructures and stoichiometries of the three samples.

With a temperature-dependent measurement of torque in the *ab*-plane at an intermediate field, we obtain the temperature-dependence of $K_{22}$ as shown in Fig. 6. The overall temperature dependence is the similar for all three samples with relatively minor differences. As temperature increases, the magnitude of $K_{22}$ decreases and becomes close to zero at $T > 150$ K. $K_{22}$ of sample A becomes slightly positive for $T > 150$ K. From Eq. (2), the total energy reaches a minimum when the Néel vector is along the crystal *a*- and *b*-axis ($\theta = 0°$ or $\pm 90°$) for $K_{22} < 0$ at zero field. When $K_{22} > 0$, the Néel vector lies in directions with $\theta = \pm 45°$ [8].

## 3. First-principles calculations of magnetocrystalline anisotropy

Magnetocrystalline anisotropy of Fe$_2$As has two contributions, one from spin-orbit interaction (SOI) and one from classical magnetic dipole-dipole interaction (MDD): anisotropy from SOI is calculated using DFT for noncollinear magnetism with spin-orbit coupling, by rotating the Néel vector both within the easy plane (001) and out of the plane towards the hard axis (010). The corresponding total-energy changes are visualized in Fig. S2 and the anisotropy energies are then obtained by fitting the energy change *vs.* Néel vector orientation to Eq. (2). This leads to a two-fold symmetric SOI anisotropy energy for the Néel vector in the (010) plane and a four-fold symmetric one for the (001) plane. In DFT-LDA, the (010) plane anisotropy energy is $K_1 = -320$ kJ/m$^3$ and $K_{22} = -290$ J/m$^3$ for the (001) plane. A non-zero $K_{22}$ indicates the existence of two local energy minima. In DFT-PBE, the corresponding values are $K_1 = $ -530 kJ/m$^3$ for the (010) plane and $K_{22} = $ 280 J/m$^3$ for the (001) plane. The sign of $K_{22}$ differing between DFT-LDA and DFT-PBE implies that the energetic ordering of these two minima is inverted. Negative $K_{22}$ means that the energy minimum occurs for a Néel vector along a <100> equivalent direction and positive $K_{22}$ for a Néel vector along a <110> equivalent direction.

The MDD contribution is computed using a classical model that is parametrized using the chemical and magnetic ground-state structure from DFT. We use the ground-state chemical and magnetic structure from DFT-LDA as well as DFT-PBE, to evaluate the following expression for the classical magnetic dipole-dipole interaction and to compare the influence of exchange and correlation:

$$E_d = -\frac{1}{2}\frac{\mu_0}{4\pi}\sum_{i \neq j}\frac{3\left[m(r_i)\cdot r_{ij}\right]\left[m(r_i)\cdot r_{ij}\right] - m(r_i)\cdot m(r_j) r_{ij}^2}{r_{ij}^5} \qquad (8)$$

To obtain the anisotropy energy for bulk Fe$_2$As from this expression, we use an interaction shell boundary $r_{ij}$ of 180 Å, which converges the result to within $10^{-7}$ eV. This leads to a two-fold symmetric MDD contribution to the anisotropy energy in the (010) plane of $K_1 = $ -220 kJ/m$^3$ for LDA. For PBE, the corresponding value is $K_1 = $ -300 kJ/m$^3$. The MDD contribution in the (001) plane is less than 1 neV and, thus, negligible.

Therefore, we find a total out-of-plane anisotropy energy of -540 kJ/m$^3$ and -830 kJ/m$^3$ from LDA and PBE, respectively. We attribute ~2/3 of the total out-of-plane anisotropy energy to the SOI contribution and ~1/3 to the MDD contribution. Both terms show two-fold symmetry with the hard axis along the <001> direction. Torque magnetometry can only measure a lower bound of $|K_1| > $ 36 kJ/m$^3$ for the out-of-plane anisotropy energy and does not contradict our DFT results.

The in-plane anisotropy energy is computed as $K_{22} = $ -290 J/m$^3$ (DFT-LDA) and $K_{22} = $ 280 J/m$^3$ (DFT-PBE), while the measured result is $K_{22} = $ -150 J/m$^3$.

## 4. Antiferromagnetic resonance of easy plane antiferromagnets

Without anisotropy, the magnon dispersion of energy in antiferromagnet is zero at the center of the Brillouin zone. Anisotropy introduces a band gap at the zone center. The antiferromagnetic resonance (AFMR) mode we refer to in this work describes this precessional magnetization motion at the zone center. With the anisotropy values we determined by theory and experiment, we can make an estimation of the AFMR frequency.

We start from equations of motion under the 'macrospin' approximation of the two magnetic sublattices in domain D1 [28][29]:

$$\frac{dM_1}{dT} = |\gamma| M_1 \times \left[ (H_{22} + H_{ex})\hat{i} + \left(-\lambda M_2^b + \frac{M_1^b}{M} H_{22}\right)\hat{j} + \left(-\lambda M_2^c + \frac{M_1^c}{M} H_1\right)\hat{k} \right] \quad (9)$$

$$\frac{dM_2}{dT} = |\gamma| M_2 \times \left[ (-H_{22} - H_{ex})\hat{i} + \left(-\lambda M_1^b + \frac{M_2^b}{M} H_{22}\right)\hat{j} + \left(-\lambda M_1^c + \frac{M_2^c}{M} H_1\right)\hat{k} \right] \quad (10)$$

Where $\gamma$ is the gyromagnetic ratio, $M_1$ and $M_2$ are sublattice magnetizations of domain D1, $H_1$ and $H_{22}$ are out-of-plane and in-plane anisotropy fields, respectively, which can be written as $H_1 = K_1/M$ and $H_{22} = K_{22}/M$. $H_{ex} = \lambda M$ and $\lambda$ is the inter-sublattice exchange interaction. $M_1^a = -M_2^a \approx M$, $M^b$ and $M^c$ are magnetization components along the *b*- and the *c*-axis, respectively, during spin procession.

Because the two sublattices along the *a*-axis are aligned antiparallel to each other, the anisotropy fields along the *a*-axis are of opposite signs. Along the *b*- and *c*-axes, the sign of the effective anisotropy field is determined by the signs of $M_{1,2}^b$ and $M_{1,2}^c$. In domain D1, although the sublattice magnetizations stay along the *a*-axis, the in-plane anisotropy is of four-fold symmetry, so there is equivalent anisotropy energy contribution along the *a*- and *b*-axes. The effective anisotropy fields along the *a*- and *b*-axes are determined by the projection of magnetization on these axes.

The only non-zero solution of the equation of motion requires $M_1^b = M_2^b$ and $M_1^c = -M_2^c$, as shown in Fig. S2. The corresponding angular frequency can be expressed as $\omega = |\gamma| \sqrt{2H_{ex}(H_{22} - H_1)}$.

In easy-plane AFs, $K_1 < 0$ and its absolute value is usually much larger than $K_{22}$, thus the frequency is always real. Besides, the AFMR frequency is smaller with smaller $H_{22} - H_1$ value, because the system is more isotropic. For easy-plane materials with $K_{22} - K_1 \ll 0$, the AFMR frequency is dominated by the anisotropy in the direction perpendicular to the easy plane.

For the exchange field, we use sublattice magnetization $M_{D1} = 4 \times 10^5$ A/m and an exchange integral $\lambda \approx 1/\chi_\perp$ with calculated $\chi_\perp = 0.0036$. We obtain an exchange field $H_{ex} \approx 140$ T. With calculated $K_1$ value from DFT-PBE, $K_1 = -830$ kJ/m$^3$, the AFMR frequency is $f = 670$ GHz.

For tetragonal antiferromagnets like Fe$_2$As, the AFMR is dominated by $K_1$ because $|K_1| \gg |K_{22}|$. The same relation is also valid for Mn$_2$Au where a previous calculation [30] shows that the magnitude of the out-of-plane anisotropy is also much larger than the in-plane anisotropy. It is important to determine $K_1$ to estimate AFMR frequency, and both $K_1$ and $K_{22}$ value are needed for thermal stability of spintronics materials.

As discussed in Refs. [2] and [31], the electrical current typically switches only a small number of antiferromagnetic domains. As the anisotropy energy scales with sample volume, the total in-plane anisotropy energy is determined by the volumetric difference of the two kinds of domains, $\Delta E = (V_{D1} - V_{D2})K_{22}$. $V_{D1} - V_{D2}$ depends on temperature, current density and the pulse width [32]. If the attempt frequency is not high enough and $V_{D1} - V_{D2}$ is small, the magnetic state is more susceptible to thermal fluctuations.

## CONCLUSION

We performed torque magnetometry measurements on an easy-plane antiferromagnet Fe$_2$As. The measurement results prove that the domain wall motion in the single-crystalline sample is reversible, and allow us to extract the in-plane anisotropy when the magnetic energy $E_m$ is comparable to magnetocrystalline anisotropy energy $E_{ani}$. The in-plane anisotropy of Fe$_2$As is $K_{22}$ = -150 J/m$^3$ at 4 K. $K_{22}$ is strongly temperature-dependent and its magnitude decreases as a function of temperature. This means that the domain structure in Fe$_2$As may be easily perturbed by a small applied field at room temperature. With $K_1$ = -830 kJ/m$^3$ calculated from DFT, we derived the AFMR frequency $f = \dfrac{|\gamma|}{2\pi}\sqrt{2H_{ex}(H_{22} - H_1)}$ = 670 GHz. Our analysis of torque magnetometry data suggests that the in-plane magnetic anisotropy of some candidate materials for antiferromagnetic spintronic applications, such as Fe$_2$As, can be very small at room temperature. A field smaller than 1 T is sufficient to significantly alter its domain structure. The measurement of $K_{22}$ in Fe$_2$As provides a baseline value for further studies of magnetic anisotropy of easy-plane antiferromagnets and the motion of antiferromagnetic domain walls.

## ACKNOWLEDGEMENTS

This work was undertaken as part of the Illinois Materials Research Science and Engineering Center, supported by the National Science Foundation MRSEC program under NSF Award No. DMR-1720633. This work made use of the Illinois Campus Cluster, a computing resource that is operated by the Illinois Campus Cluster Program (ICCP) in conjunction with the National Center for Supercomputing Applications (NCSA) and which is supported by funds from the University of Illinois at Urbana-Champaign. This research is part of the Blue Waters sustained-petascale computing project, which is supported by the National Science Foundation (Awards No. OCI-0725070 and No. ACI-1238993) and the state of Illinois. Blue Waters is a joint effort of the University of Illinois at Urbana-Champaign and its National Center for Supercomputing

Applications. Z.D. acknowledges support from the Swedish Research Council (VR) under Grant No. 2015-00585, co-funded by Marie Skłodowska-Curie Actions (Project INCA 600398).

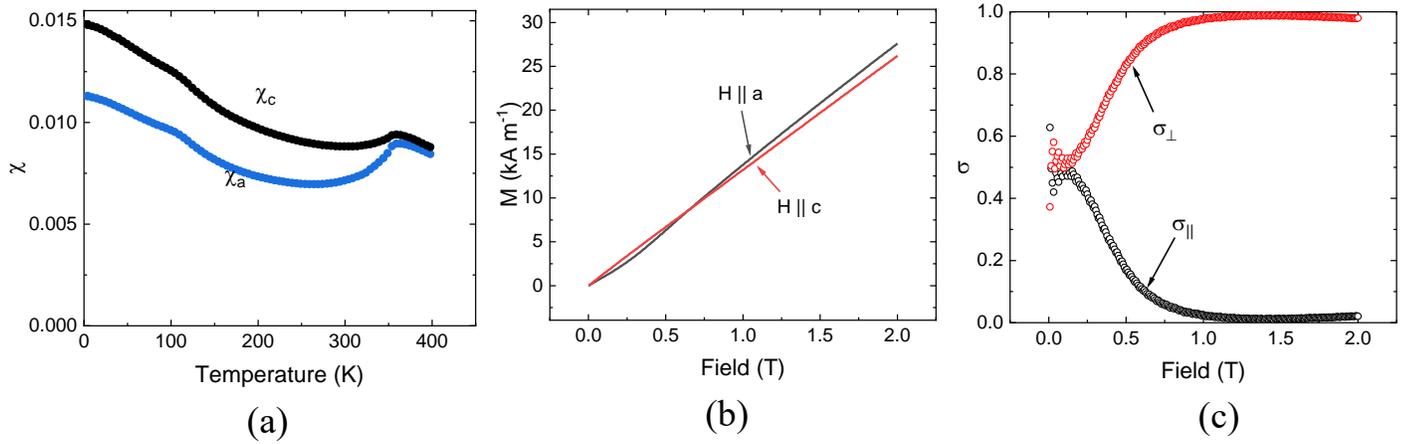

Figure 1. (a) Temperature dependence of the magnetic susceptibility of Fe$_2$As in the low field limit as measured using 10 mT field applied along the *a*-axis (blue data points) and *c*-axis (black data points) of the crystal. (b) Dependence of Fe$_2$As magnetization *M* on applied field *H* at *T* = 4 K. With *H* along the *c*-axis (red line), *M* is a linear function of *H*. With *H* along the *a*-axis (black curve), the non-linear dependence of *M* on *H* is due to the rotation of antiferromagnetic domains. (c) The population of domains with Néel vectors parallel and perpendicular to the applied field estimated from the dependence of *M* on *H*. The assumptions are: 1) in zero field, the population of domains with Néel vectors in the *a* and *b* directions are equal; and 2) in the high field limit, the Néel vector is perpendicular to the applied field.

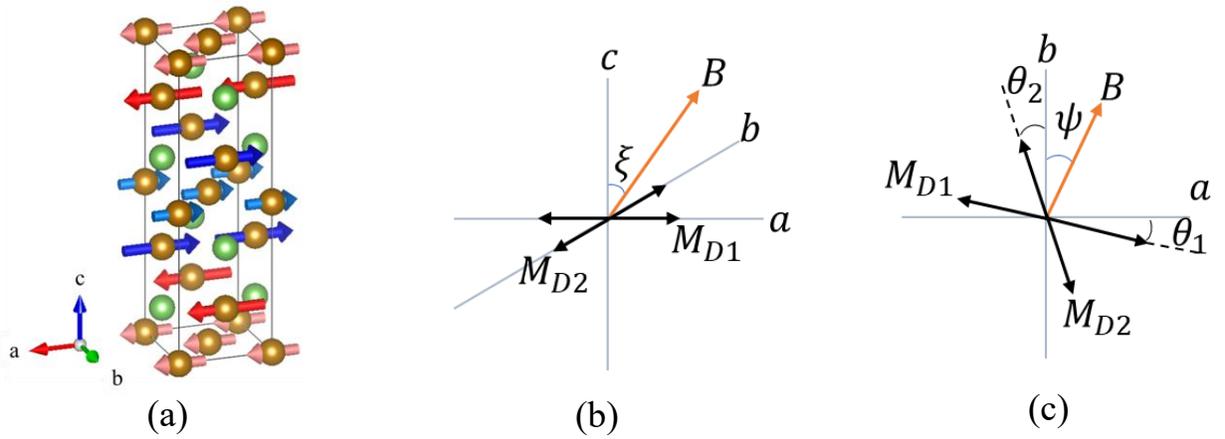

Figure 2. Geometry of the torque magnetometry experiments. The *a-b-c* coordinates are the crystal axes. $M_{D1}$ and $M_{D2}$ are the sublattice magnetizations of the two types of domains labeled as D1 and D2. (a) The magnetic unit cell of $Fe_2As$. (b) Three-dimensional perspective of the measurement with the magnetic field rotating in the *ac*-plane. The magnetic field makes an angle $\xi$ with the *c*-axis of the crystal. $M_{D1}$ and $M_{D2}$ are assumed to stay along the *a*- and the *b*-axis, respectively. The torque is along the *b*-axis. (c) Plan-view of the measurement with the magnetic field rotating in the *ab*-plane. The magnetic field makes an angle $\psi$ with the *b*-axis of the crystal. $M_{D1}$ and $M_{D2}$ tilt away from *a*- and *b*-axis by $\theta_1$ and $\theta_2$, respectively ($\theta_1$ and $\theta_2$ are not necessarily the same). The torque is along the *c*-axis (normal to the plane of the drawing).

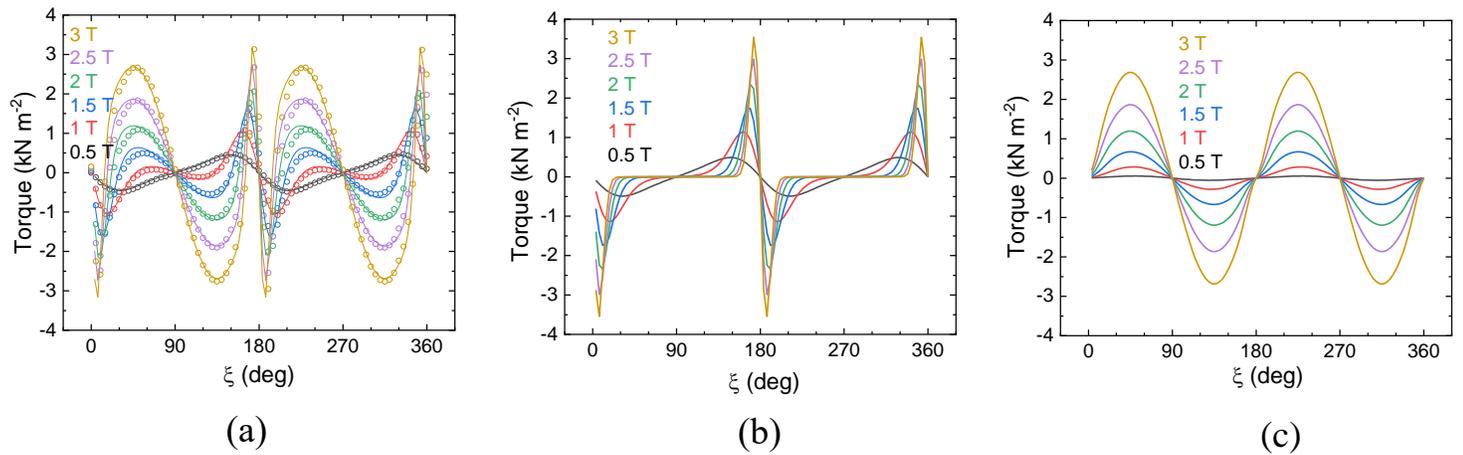

Figure 3. (a) Torque magnetometry measurements in the *ac*-plane of Fe$_2$As at $T$ = 4 K. Open symbols are measured data; solid lines are fits to the data. The legend gives the magnitude of the applied field labeled by color. (b) Calculated torque generated by domains of type D1 as a function of applied field. (c) Calculated torque generated by domains of type D2 as a function of applied field.

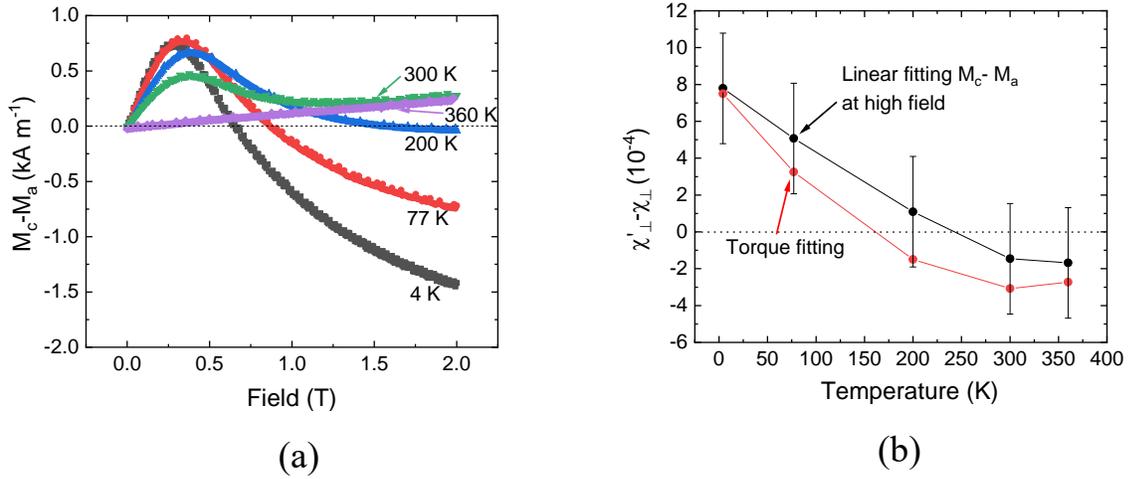

Figure 4. (a) The field dependence of the difference $M_c - M_a$ between the magnetization with an applied field along the $c$ direction $M_c$ and the field applied along the a direction $M_a$. Each curve is labeled by the measurement temperature. When the applied field along the $a$-axis is larger than 1.5 T, all domains can be treated as equivalent to $M_{D2}$. Therefore, the slope of the data for magnetic fields larger than 1.5 T is the difference in the susceptibility $\chi_\perp - \chi'_\perp$, where $\chi'_\perp$ is the susceptibility in ab-plane perpendicular to $M_{D2}$ and $\chi_\perp$ is the susceptibility along $c$-axis perpendicular to $M_{D2}$. (b) Comparison of the temperature dependent of $\chi'_\perp - \chi_\perp$ value from direct measurements of the type shown in panel (a) and from fitting the torque data.

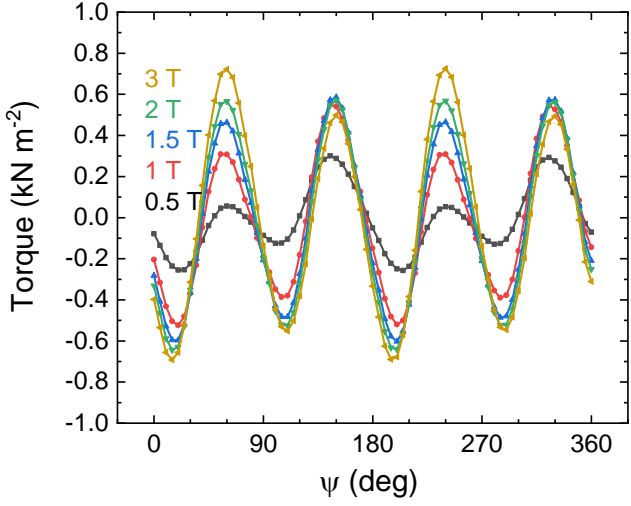 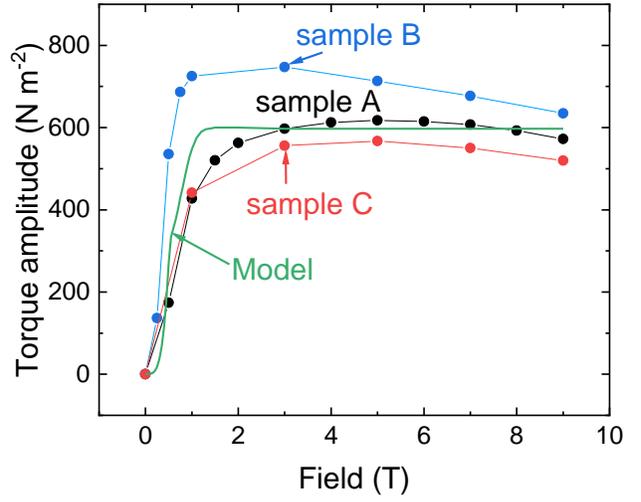

(a) (b)

Figure 5. (a) Torque magnetometry measurements of Fe$_2$As (sample A) in the *ab*-plane at $T$=4 K. (b) The amplitude of the four-fold component of the torque extracted from measurements of the type shown in panel (a) at $T = 4$ K and compared to an analytical model (see text). When amplitude of the four-fold component of the torque saturates at the value $\tau_0$, the in-plane anisotropy is $\tau_0 \approx 4K_{22}$.

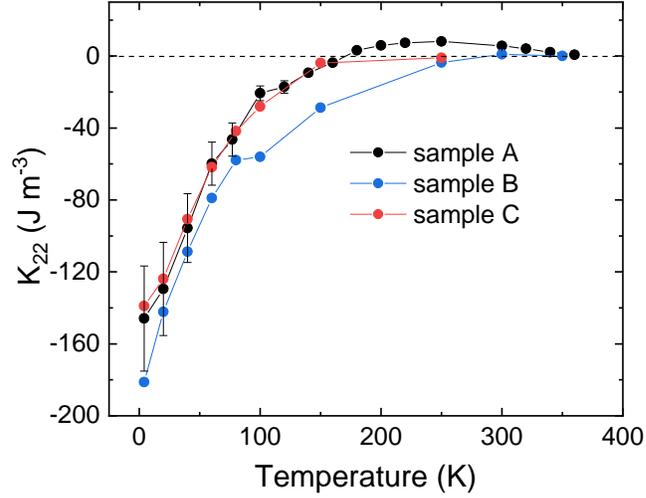

Figure 6. Temperature dependence of in-plane magnetocrystalline anisotropy $K_{22}$ of $Fe_2As$. The torque data with the external field rotating in the (001) plane were measured with an intermediate field strength (3 T for sample A and sample C, and 1 T for sample B) and the amplitude of the four-fold symmetry was extracted to obtain the in-plane anisotropy with $\tau_0 \approx 4K_{22}$. Intermediate field is defined as a field strength under which the torque amplitude of four-fold symmetry saturates at 4 K. The error bars represent 20% uncertainty in determining the saturation value of the torque amplitude. All three samples have the same uncertainty, and we only plot error bar for sample A to ensure that the plot is free from cluttering.